\documentclass[notitlepage,twocolumn,prd,preprintnumbers,longbibliography,showpacs,10pt]{revtex4-2}
\usepackage[colorlinks=true,urlcolor=blue,linkcolor=blue,citecolor=blue]{hyperref}
\usepackage{slashed,color,amsmath,amssymb,mathrsfs}
\usepackage{amsfonts,epsfig,amssymb}
\usepackage{graphicx}
\usepackage{hyperref}
\usepackage{longtable}
\usepackage{array}
\usepackage{accents}
\usepackage{tensor}
\usepackage{footmisc}
\usepackage{longtable}
\usepackage{booktabs}
\usepackage{multirow}
\usepackage{bm}
\usepackage{subfigure}
\usepackage{gensymb}
\usepackage{booktabs}

\newcommand{\dd}{\mathrm{d}}
\newcommand{\mjp}{m_{J/\psi}}
\newcommand{\Jp}{J/\psi}

\allowdisplaybreaks

\begin{document}

\title[]{Structure around  \texorpdfstring{$p\bar{p}$}{} threshold in $J/\psi$ radiative decays}
\author{Qin-He Yang$^{1,2}$}
\author{Di Guo$^{1,2}$}
\author{Ling-Yun Dai$^{1,2}$}
\email{dailingyun@hnu.edu.cn}
\affiliation{$^{1}$ School of Physics and Electronics, Hunan University, Changsha 410082, China}
\affiliation{$^{2}$ Hunan Provincial Key Laboratory of High-Energy Scale Physics and Applications, Hunan University, Changsha 410082, China}

\date{\today}
\begin{abstract}
In this paper, we study the structure around the $p\bar{p}$ threshold that appears in $\eta'\pi^+\pi^-$, $3(\pi^+\pi^-)$ and $K_S^0K_S^0\eta$ invariant mass spectra in the processes of relevant $J/\psi$ radiative decays. The $N\bar{N}$ rescattering is taken into account, and the distorted-wave Born approximation is applied to get the decaying amplitude through a two-step process: $J/\psi\to\gamma N\bar{N}\to\gamma\eta'\pi^+\pi^-$, $\gamma 3(\pi^+\pi^-)$ and $\gamma K_S^0K_S^0\eta$. The $N\bar{N}$ scattering amplitudes are obtained by solving the Lippmann-Schwinger equation with the potentials given by chiral effective field theory. To fix the unknown couplings, we fit the amplitudes to the datasets of the latest measurements on the invariant mass spectra of $J/\psi$ radiative decays, as well as the phase shifts and inelasticities given by partial wave analysis. We vary the cutoffs ($R$=0.9, 1.0,  and 1.1 fm) and find that the solutions are stable. The structures around $p\bar{p}$ threshold found in the processes of $J/\psi\to\gamma\eta'\pi^+\pi^-$, $J/\psi\to\gamma3(\pi^+\pi^-)$ and $J/\psi\to\gamma K_S^0K_S^0\eta$ can be attributed to threshold behavior of $N\bar{N}$ intermediate states.
\end{abstract}
\maketitle

\section{Introduction}\label{Sec:I}
Physicists have long been interested in 6-quark dibaryons and 3-quark-3-antiquark baryonium states. They can be one new kind of inner structure of matters and give clues to experimentalists to find new resonances. 
One possible candidate of the baryonium, $X(1835)$ first discovered in the invariant mass spectrum of $M_{\eta'\pi^+\pi^-}$ near the $p\bar{p}$ threshold in the process of $\Jp\to\gamma\eta'\pi^+\pi^-$ by BES collaboration \cite{BES:2005ega}, attracts both theoretical and experimental attention as the proton and neutron are the basic components of nucleus.  Its mass and width are given as $M=1833.7\pm 6.1$ MeV$/c^2$ and $\Gamma=67.7\pm 20.3$ MeV$/c^2$ \cite{BES:2005ega}. Subsequently, this resonance is also observed in decays of $\Jp\to\gamma 3(\pi^+\pi^-)$ \cite{BESIII:2013sbm}, $\Jp\to\gamma K_S^0 K_S^0 \eta$~\cite{BESIII:2015xco} and $\Jp\to\gamma\phi$~\cite{BESIII:2018dim}, while it is faintly supported by a few other experiments \cite{Belle:2013jng,Belle:2012uhr}. 
A few years ago, BESIII increased the statistics of their measurements on the $\Jp\to\gamma\eta'\pi^+\pi^-$ and observed a clear structure around $p\bar{p}$ threshold, but it does not look like a Breit-Wigner type peak around 1835~MeV.  
Indeed, the anomalous behavior around $p\bar{p}$ threshold has also been observed in quite a bit other processes such as $B^{\pm}\to K^{\pm}p\bar{p}$ and $\bar{B}^0\to D^{*0}p\bar{p}$ by Belle collaboration~\cite{Belle:2002bro,Belle:2002fay}, $\Jp\to\gamma p\bar{p}$ by BES collaboration~\cite{BES:2003aic}, $e^+e^-\to p\bar{p}/n\bar{n}$ by BABAR \cite{BaBar:2013ves} and SND \cite{SND:2022wdb} collaborations, and $e^+e^-\to 2(\pi^+\pi^-\pi^0), 3(\pi^+\pi^-)$ by BABAR collaboration \cite{BaBar:2006vzy}. 
Consequently, one would wonder about the internal links between the structure around $p\bar{p}$ threshold in the energy region around the $X(1835)$ and $N\bar{N}$ rescattering.  
There are many relevant theoretical researches which focused on $e^+e^-\to p\bar{p}/n\bar{n}$~\cite{Haidenbauer:2014kja,Yang:2022qoy}, $\Jp\to \gamma p\bar{p}$~\cite{Kang:2015yka,Dai:2018tlc,Dedonder:2018ulo}, and $e^+e^-\to 2(\pi^+\pi^-\pi^0), 3(\pi^+\pi^-)$ \cite{Haidenbauer:2015yka}, etc.. 

In the previous paper \cite{Dai:2018tlc}, the $N\bar{N}$ final-state interactions (FSI)\footnote{For different approaches implementing FSI, see e.g. Refs.\cite{Dai:2012pb,Dai:2014zta,Yao:2020bxx}. } have been included to study the structure around $p\bar{p}$ threshold in the process of $\Jp\to\gamma\eta'\pi^+\pi^-$. 
The strategy is following a two-step process, $\Jp\to\gamma N\bar{N}\to\gamma \eta'\pi^+\pi^-$, with the quantum number of $N\bar{N}$ being $^1S_0$. The distorted-wave Born approximation (DWBA) is used to implement FSI of $N\bar{N}$. The underlying physics behind the first process $\Jp\to\gamma N\bar{N}$ is following the experimental measurements of anomalous energy-dependent behavior near the $p\bar{p}$ threshold in $\Jp\to\gamma N\bar{N}$. See, e.g., Ref.~\cite{BES:2003aic}, where there is an apparent threshold enhancement near the $p\bar{p}$ threshold in the decay of $\Jp\to\gamma p\bar{p}$ as observed by BES collaboration.   
Echoing these measurements, some theoretical models do find that the structure around $p\bar{p}$ threshold discovered in $\eta'\pi^+\pi^-$ invariant mass spectrum of $\Jp\to\gamma\eta'\pi^+\pi^-$ and/or the cross section of electron-positron annihilation into multipions is caused by the intermediate $N\bar{N}$ rescattering~\cite{Kang:2015yka,Dai:2018tlc}, while no resonance pole is found in the $^1S_0$ partial wave. 

Recently, BESIII collaboration performed the latest measurements on $\Jp\to\gamma\eta'\pi^+\pi^-$ again, and the uncertainty of the $\eta'\pi^+\pi^-$ invariant mass spectrum has been reduced near the $p\bar{p}$ threshold ~\cite{Ablikim2022}. Thus, it would be necessary to include these new datasets and update the research of Ref.~\cite{Dai:2018tlc}. 
Also, there are clear structures around $p\bar{p}$ threshold in the measurements of the processes of $\Jp\to\gamma3(\pi^+\pi^-)$ and $\Jp\to\gamma K_S^0K_S^0\eta$. 
If the structure is caused by the $N\bar{N}$ threshold behavior, the DWBA method discussed above should give a compatible data description. Thus we will include all these processes and give a systematic discussion. 
Further, in the previous work, only one cutoff (R=0.9~fm) was considered, and it is essential to check whether the structure is stable with different cutoffs. 

This paper is organized as follows. In  Sec.~\ref{Sec:II}, the formulas to calculate the reactions of $\Jp\to\gamma\eta'\pi^+\pi^-$, $\Jp\to\gamma3(\pi^+\pi^-)$, $\Jp\to\gamma K_S^0K_S^0\eta$, $\Jp\to\gamma p\bar{p}$, $p\bar{p}\to \eta'\pi^+\pi^-$, $p\bar{p}\to 3(\pi^+\pi^-)$ and $p\bar{p}\to K_S^0K_S^0\eta$ are given. Through solving a set of coupled channel equations, e.g., Lippmann-Schwinger equation (LSE) and DWBA equations, one can get the scattering and decaying amplitudes.  
Then the fit results and discussions are given in Sec.~\ref{Sec:III}. The $\Jp\to\gamma p\bar{p}$ amplitude is fixed by fitting to the $p\bar{p}$ invariant mass spectra and the phase shifts of $^1S_0$ partial waves of $N\bar{N}$ scattering. The amplitudes of $p\bar{p}$ scattering into $ \eta'\pi^+\pi^-$, $3(\pi^+\pi^-)$ and $K_S^0K_S^0\eta$ are fixed by fitting to the invariant mass spectra and/or branching ratios. 
The $\Jp\to\gamma\eta'\pi^+\pi^-$, $\Jp\to\gamma3(\pi^+\pi^-)$ and $\Jp\to\gamma K_S^0K_S^0\eta$ amplitude is then fixed by fitting to the corresponding invariant mass spectra of $J/\psi$ radiative decays. The effects of varying cutoffs are also discussed. 
Finally, a summary is given in Sec.~\ref{Sec:IV}.

\section{Formalism}\label{Sec:II}
As mentioned above, the $\Jp\to\gamma \eta'\pi^+\pi^-$, $\Jp\to\gamma 3(\pi^+\pi^-)$ and $\Jp\to\gamma K_S^0 K_S^0\eta$ decaying amplitudes are obtained through two-step processes of $\Jp\to\gamma N\bar{N} \to \gamma\eta'\pi^+\pi^-$, $\gamma 3(\pi^+\pi^-)$ and $\gamma K_S^0 K_S^0\eta$. The $N\bar{N}$ off-shell scattering amplitude is the kernel to be input into DWBA equations, which is solved by LSE \cite{Dai:2017ont}, 
\begin{eqnarray}\label{eq:LSE}
    T_{N\bar{N}\to N\bar{N}}&=&V_{N\bar{N}\to N\bar{N}}+V_{N\bar{N}\to N\bar{N}}G_0T_{N \bar{N}\to N\bar{N}}\,,
\end{eqnarray}
where $T_{N\bar{N}\to N\bar{N}}$ is the $N\bar{N}$ scattering amplitude. Here the relevant partial waves are of $^1S_0$, with isospin $I=0,1$. $G_0$ denotes the free propagator of the nucleon, $V_{N\bar{N}\to N\bar{N}}$ is the nucleon-antinucleon interaction potential, which is calculated from chiral effective field theory (ChEFT) up to next-to-next-to-next-to leading order (N$^3$LO). See Ref.~\cite{Dai:2017ont} for details.  
The other amplitudes of the processes of $\Jp\to\gamma N\bar{N}$, $N\bar{N} \to \nu$ and $\Jp\to\gamma \nu$, are obtained by solving the following set of coupled equations established by DWBA, 
\begin{eqnarray}\label{eq:DWBA}
    F_{N\bar{N}\to\nu}&=&V_{N\bar{N}\to\nu}+T_{N\bar{N}\to N\bar{N}}G_0V_{N\bar{N}\to\nu}\,,\nonumber\\
    F_{\Jp\to\gamma N\bar{N}}&=&A^0_{\Jp\to\gamma N\bar{N}}+A^0_{\Jp\to\gamma N\bar{N}}G_0T_{N\bar{N}\to N\bar{N}}\,,\nonumber\\
    F_{\Jp\to\gamma\nu}&=&A^0_{\Jp\to\gamma\nu}+F_{\Jp\to\gamma N\bar{N}}G_0V_{N\bar{N}\to\nu}\,,
\end{eqnarray}
where $\nu$ denotes the final states with $\eta'\pi^+\pi^-$, $3(\pi^+\pi^-)$, and $ K_S^0 K_S^0\eta$, respectively.
The transition amplitudes $A^0_{\Jp\to\gamma N\bar{N}}$ and $A^0_{\Jp\to\gamma\nu}$, and the annihilation potential $V_{N\bar{N}\to\nu}$  are parameterized as
\begin{eqnarray}
    A^0_{\Jp\to\gamma N\bar{N}}(p)&=&\tilde{C}_{\Jp\to\gamma N\bar{N}}+C_{\Jp\to\gamma N\bar{N}}p^2\,,\nonumber\\
    V_{N\bar{N}\to\nu}(p)&=&\tilde{C}_{N\bar{N}\to\nu}+C_{N \bar{N}\to\nu}p^2\,,\nonumber\\
    A^0_{\Jp\to\gamma\nu}(Q)&=&\tilde{C}_{\Jp\to\gamma\nu}+C_{\Jp\to\gamma\nu}Q\,,  \label{AV}
\end{eqnarray}
where $p$ is the center-of-mass momentum of the nucleon-antinucleon system, and $Q$ is the invariant mass $M_{\eta'\pi^+\pi^-}$, $M_{3(\pi^+\pi^-)}$ and $M_{K_S^0 K_S^0 \eta}$ for different processes, respectively. 
Notice that for the processes of $\Jp\to\gamma 3(\pi^+\pi^-)$ and $p \bar{p}\to 3(\pi^+\pi^-)$, we consider $\pi^+ \pi^-$ as a whole particle with the mass $2 m_{\pi}$. Then these multipion final states are simplified into \lq\lq three body" final states and have similar formalism as other processes. The decay rates and cross sections of these processes can be obtained through
\begin{eqnarray}
    \frac{\dd\Gamma}{\dd Q}&=&\int_{t_1^-}^{t_1^+}\dd t_1\int_{t_2^-}^{t_2^+}\dd t_2\frac{(\mjp^2-Q^2)|\mathcal{M}_{\Jp\to\gamma\nu}|^2}{6144 \tilde{N} \pi^5\mjp^3 Q}\,, \nonumber\\
    \frac{\dd \Gamma}{\dd Q}&=&\frac{\lambda^{1/2}(\mjp^2,Q^2,m_{p}^2)\sqrt{Q^2-4m_{p}^2}}{2^7\pi^3\mjp^3}|\mathcal{M}_{\Jp\to\gamma p\bar{p}}|^2\,,\nonumber\\
    \sigma(Q)&=&\int_{t_1^-}^{t_1^+}\dd t_1 \int_{t_2^-}^{t_2^+}\dd t_2\frac{|\mathcal{M}_{p\bar{p}\to\nu}|^2}{1024 \tilde{N} \pi^3Q^3\sqrt{Q^2-4m_p^2}}\,,    \label{eq:dGdQ}
\end{eqnarray}
where $\mathcal{M}_{\Jp\to\gamma\nu}$, $\mathcal{M}_{\Jp\to\gamma\bar{p}p}$ and $\mathcal{M}_{\bar{p}p\to\nu}$ are the Lorentz invariant amplitudes of corresponding progresses. $\tilde{N}$ is the normalization factor caused by the property of identical particles, with $\tilde{N}=1,2,6$ for final states of $\eta'\pi^+\pi^-$, $K_S^0 K_S^0 \eta$ and $3(\pi^+\pi^-)$, respectively. One has
 \begin{eqnarray}
\!&\mathcal{M}_{\Jp\to\gamma\nu}& \!=\!-32\pi^{\frac{7}{2}}\sqrt{E_\gamma E_{\Jp}E_{1}E_{2}E_{3}}F_{\Jp\to\gamma\nu} \,, \nonumber\\   
    \!&\mathcal{M}_{\Jp\to\gamma p\bar{p}}&\!=-8\pi^2E_N\sqrt{E_\gamma E_{\Jp}}F_{\Jp\to\gamma p\bar{p}}\!\,,\nonumber\\
    \!&\mathcal{M}_{N\bar{N}\to\nu}&\!=\!-32\pi^{\frac{7}{2}}E_N\sqrt{E_{1}E_{2}E_{3}}F_{N\bar{N}\to\nu}\!\, . \label{Eq:M;inv}
\end{eqnarray}
where one has $E_{1,2,3}=E_{\eta',\pi^+,\pi^-}$, $E_{1,2,3}=E_{\eta,K_S^0,K_S^0}$ and $E_{1,2,3}=E_{(\pi^+\pi^-),(\pi^+\pi^-),(\pi^+\pi^-)}$ for different processes, respectively.
  
Notice that the meaning of invariant mass $Q$ has been extended: it can be either $Q=M_{\nu}$ for the processes of $\Jp\to\gamma\nu$ and $N\bar{N}\to\nu$ or $Q=M_{p\bar{p}}$ for $\Jp\to\gamma p\bar{p}$. 
The relations between the energies and the invariant mass $Q$ are given as 
\begin{eqnarray}
    E_{\Jp}&=&\frac{\mjp^2+Q^2}{2Q}\,,\nonumber\\
    E_\gamma&=&\frac{\mjp^2-Q^2}{2Q}\,,\nonumber\\
    E_{1}&=&\frac{Q^2-t_1+m_1^2}{2Q}\,,\nonumber\\
    E_{2}&=&\frac{Q^2-t_2+m_2^2}{2Q}\,,\nonumber\\
    E_{3}&=&\frac{t_1+t_2-m_2^2-m_1^2}{2Q}.
\end{eqnarray}
where the subscripts \lq 1,2,3' have the same meaning as those appear in Eq.(\ref{Eq:M;inv}).
The upper and lower limits of the integrals of Eq.(\ref{eq:dGdQ}) are
\begin{eqnarray}
    t_1^-&=&4 m_2^2\,,\nonumber\\
    t_1^+&=&(Q-m_1)^2\,,\nonumber\\
    t_2^-&=&\frac{1}{4t_1}\left[(Q^2-m_1^2)^2-(\lambda^{1/2}(Q^2,t_1,m_1^2)\right.\nonumber\\
    &&\left.+\lambda^{1/2}(t_1,m_2^2,m_2^2))^2\right]\,,\nonumber\\
    t_2^+&=&\frac{1}{4t_1}\left[(Q^2-m_1^2)^2-(\lambda^{1/2}(Q^2,t_1,m_1^2)\right.\nonumber\\
    &&\left.-\lambda^{1/2}(t_1,m_2^2,m_2^2))^2\right]\,, \label{Eq:tlimit}
\end{eqnarray}
with the K\"all\'en function defined as $\lambda(x,y,z)=(x-y-z)^2-4yz$. 
The explicit set of coupled equations to solve the scattering and decaying amplitudes are as follows
\begin{eqnarray}
    & &T_{N\bar{N}\to N\bar{N}}(p',p;E_p)=V_{N\bar{N}\to N\bar{N}}(p',p)+\int_0^\infty \frac{\dd k k^2}{(2\pi)^3}\nonumber\\
    & &\;\;\; V_{N\bar{N}\to N\bar{N}}(p',k)\frac{1}{2E_p-2E_k+i\epsilon}T_{N\bar{N}\to N\bar{N}}(k,p;E_p) \,, \nonumber\\
    & &F_{\Jp\to\gamma N\bar{N}}(Q)=A^0_{\Jp\to\gamma N\bar{N}}(p)+\int_0^\infty\frac{\dd k k^2}{(2\pi)^3}\nonumber\\
    & &\;\;\; A^0_{\Jp\to\gamma N\bar{N}}(k)\frac{1}{Q-2E_k+i\epsilon}T_{N\bar{N}\to N\bar{N}}(k,p;E_p) \,,\nonumber\\
    & &F_{N\bar{N}\to\nu}(Q)=V_{N\bar{N}\to\nu}(p)+\int_0^\infty\frac{\dd k k^2}{(2\pi)^3}\nonumber\\
    &&\;\;\; T_{N\bar{N}\to N\bar{N}}(p,k;E_k)\frac{1}{2E_k-Q+i\epsilon}V_{N\bar{N}\to\nu}(k)\,,\nonumber\\
    & &F_{\Jp\to\gamma\nu}(Q)=A^0_{\Jp\to\gamma\nu}(Q)+\int_0^\infty\frac{\dd k k^2}{(2\pi)^3}\nonumber\\
    & &\;\;\; F_{\Jp\to\gamma N\bar{N}}(E_k)\frac{1}{Q-2E_k+i\epsilon}V_{N\bar{N}\to\nu}(k)\,,\label{Eq:TF;all}
\end{eqnarray} 
where $E_p,E_k$ are the energy of the nucleon or antinucleon in the center of mass frame, with $E_p=Q/2$ for the processes of $\Jp\to\gamma N\bar{N}$ and $N\bar{N}\to\nu$. 
The transition potentials given in Eq.(\ref{AV}) are multiplied with an exponential regulator in the actual calculations, 
which is consistent with the regularization of $N\bar{N}$ potentials, see Ref.\cite{Dai:2018tlc} for details. This also ensures that the integrations in Eq.(\ref{Eq:TF;all}) are converging.

\section{Results and discussions}\label{Sec:III}
In this section, we will discuss the fit to the experimental data and the physics underlying the structure around $p\bar{p}$ threshold. 
Since the decaying amplitude of $\Jp\to\gamma\nu$ is constructed based on a two-step process, it would be convenient to divide this section into four parts: one is about the physics of the process of $\Jp\to\gamma p\bar{p}$;  The second part is of $\Jp\to\gamma\eta'\pi^+\pi^-$; The third part is about the results of $\Jp\to\gamma 3(\pi^+\pi^-)$; And the last part is about $\Jp\to\gamma K_S^0 K_S^0\eta$.

\subsection{Analysis of \texorpdfstring{$J/\psi\to \gamma p\bar{p}$}{}}
First, we focus on the process of  $\Jp\to\gamma p\bar{p}$. As discussed above, the $N\bar{N}$ final state interactions should be considered. It is reasonable to assume that the relative angular momentum between $\gamma$ and $p\bar{p}$ should be the lowest one, i.e., S-wave. Then the relevant partial wave of $N\bar{N}$ should be $^1S_0$, with isospin to be either one or zero according to the unfixed isospin of the photon. Here we follow the previous work~\cite{Dai:2018tlc} and set the ratio between different isospin components to be $T_{N\bar{N}\to p\bar{p}}=0.4T^{0}+0.6T^{1}$. Indeed, this ratio gives the best $\chi^2_{d.o.f.}$ for fitting to the data. 
In the previous work \cite{Dai:2018tlc}, the process of $\Jp\to\gamma p\bar{p}$ has been analyzed at $R=0.9$ fm. To study the stability of the solutions with different cutoffs, we analyze the process again with a set of cutoffs, e.g., $R=$0.9, 1.0, and 1.1 fm. 
Here the corresponding cutoffs in the momentum space are given by $\Lambda=2R^{-1}$, 
see Ref.~\cite{Dai:2017ont}. 

\begin{figure}[htp]
    \centering
    \includegraphics[width=0.99\linewidth, height=0.15\textheight]{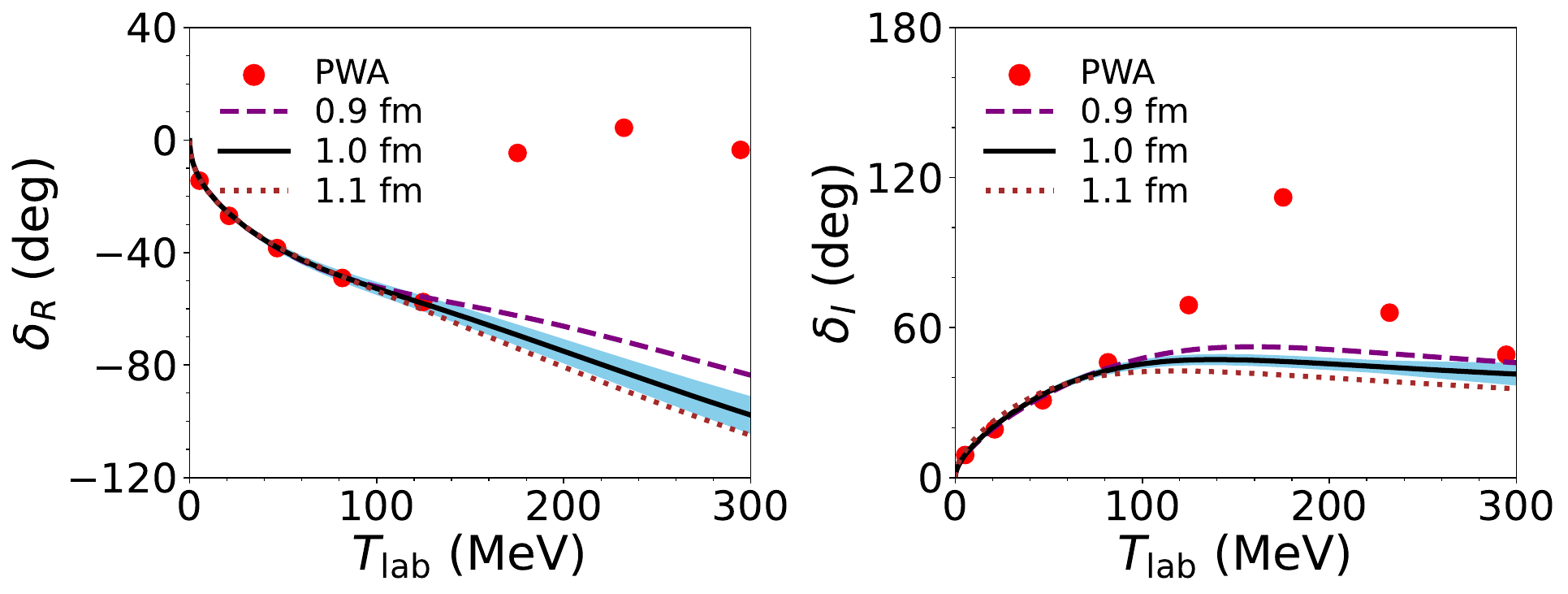}
    \includegraphics[width=0.99\linewidth, height=0.15\textheight]{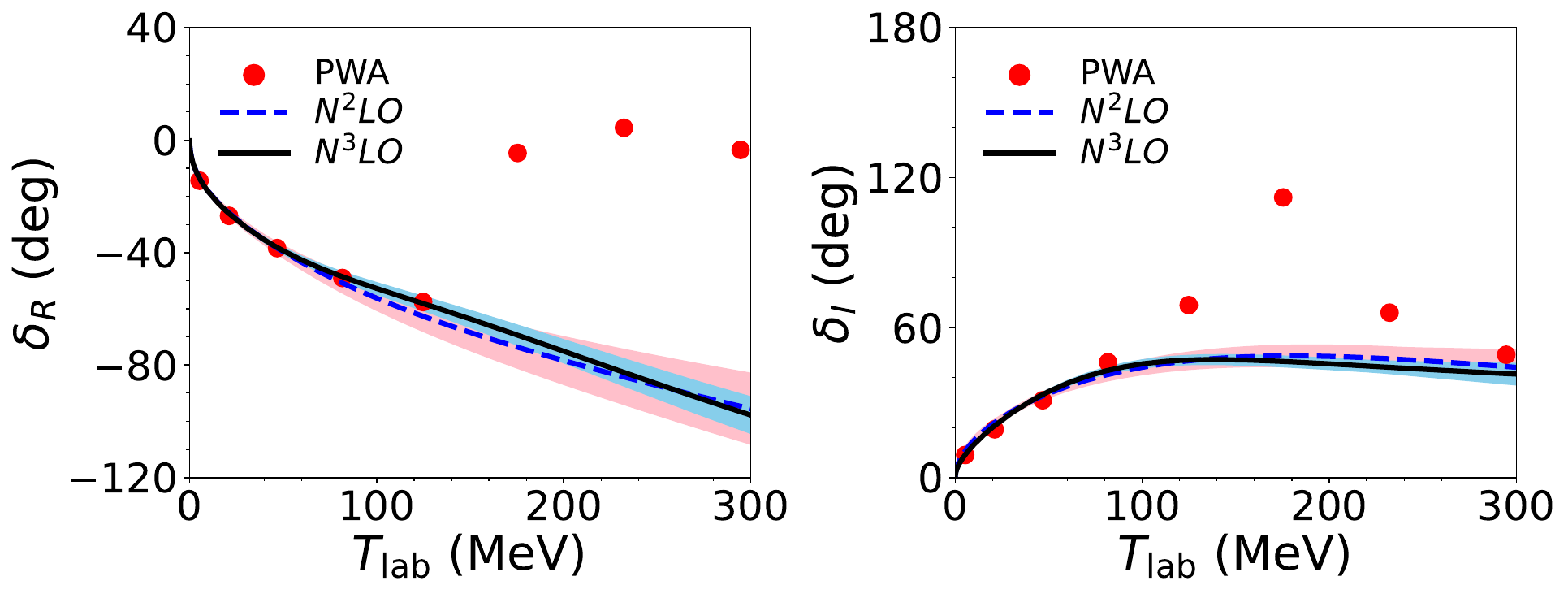}    
    \caption{ Fit to the PWA's. The top graphs are the $I=$1 phase shifts of $^1S_0$ partial wave of N$^3$LO with cutoffs $R=0.9$, 1.0 and 1.1 fm. The red points represent the results of PWA \cite{Zhou:2012ui}. The bottom graphs are the results for N$^2$LO and N$^3$LO, with cutoff $R=1.0$~fm. The pink and sky blue bands are their uncertainties, respectively. }
    \label{PSI1}
\end{figure}

The $N\bar{N}$ scattering amplitudes are solved by LSE. Similar to Ref.\cite{Dai:2018tlc},
the phase shifts of $^1S_0$ $N\bar{N}$ partial wave with isospin $I$=1 are refit to reproduce the invariant mass spectra of $\Jp\to\gamma p\bar{p}$. 
The values of the low-energy constants (LECs) of the chiral $N\bar{N}$ scattering potential of $I=1$ $^1S_0$ partial wave are listed in Table.~\ref{TlabLEC}.
\begin{table}[htp]
    \centering
    \begin{tabular}{c|c|c c c}
    \hline\hline
         LECs& N$^2$LO& \multicolumn{3}{c}{N$^3$LO}  \\
         \hline
         $R$~(fm)& 1.0 & 0.9 & 1.0 & 1.1\\\hline
         $\tilde{C}_{3^1S_0}$(GeV$^{-2}$)  & 0.0480  & 0.3535& 0.2674 & 0.0664\\
         $C_{3^1S_0}$(GeV$^{-4}$)          &-1.8397  &-3.7566&-5.1411 &-5.3367\\
         $D^1_{3^1S_0}$(GeV$^{-6}$)        & $\dots$ &-7.9227&-15.0857&-25.0000\\
         $D^2_{3^1S_0}$(GeV$^{-6}$)        & $\dots$ &11.1314& 20.0000& 25.0000\\
         $\tilde{C}^a_{3^1S_0}$(GeV$^{-1}$)&-0.0102  &-0.0120&-0.0907 &-0.1154\\
         $C^a_{3^1S_0}$(GeV$^{-3}$)        &-4.5259  &-3.0480&-3.8174 &-4.6828\\\hline\hline
    \end{tabular}
    \caption{The values of low energy constants of the $N\bar{N}$ scattering potentials of $I=1$ $^1S_0$ partial wave. All parameters are multiplied with a factor $10^4$.}
    \label{TlabLEC}
\end{table}
The fitting results of $N\bar{N}$ scattering phase shifts and inelasticity at N$^3$LO for different cutoffs are at the top of Fig.~\ref{PSI1}. 
The results of N$^2$LO and N$^3$LO with cutoff $R=1$~fm are at the bottom of Fig.~\ref{PSI1}. Both are consistent with the phase shifts and inelasticity of partial wave analysis (PWA) \cite{Zhou:2012ui} in the low-energy region ($T_{\mathrm{lab}}$ below 100 MeV). 
This is guaranteed by the fact that we fit the first four points to constraint the low energy behavior of the off-shell $N\bar{N}$ scattering amplitudes $T_{N\bar{N}\to N\bar{N}}$, which will be input in the DWBA. 
While in the high energy region, there is an apparent difference between ours and PWA's. The reason is that the phase shifts are the same in the low energy region for the processes of $\Jp\gamma\to N\bar{N}$ and $N\bar{N}\to N\bar{N}$, but they do not need to be the same in the high energy region. This is required by the final state interactions, e.g., the Watson theorem. See Refs.~\cite{Au:1986vs,Dai:2014zta,Yao:2020bxx} for more discussions.
Also, the $p\bar{p}$ invariant mass spectra of $\Jp\to \gamma p\bar{p}$ are smooth and decrease slowly in the high energy region, which requires the phase shifts to have similar behavior. 
For nucleon-antinucleon system, the relation between $T_{\mathrm{lab}}$, $p_{\mathrm{lab}}$ and the nucleon's energy, $E_N$ in the center of mass frame, is given as follows:
\begin{eqnarray}
    T_{\mathrm{lab}}=\frac{2(E_N^2-m_N^2)}{m_N}  ,\,\;\; p_{\mathrm{lab}}=\frac{2E_N\sqrt{E_N^2-m_N^2}}{m_N} \,. \nonumber
\end{eqnarray} 

\begin{figure}[htp]
    \centering
    \includegraphics[width=0.9\linewidth]{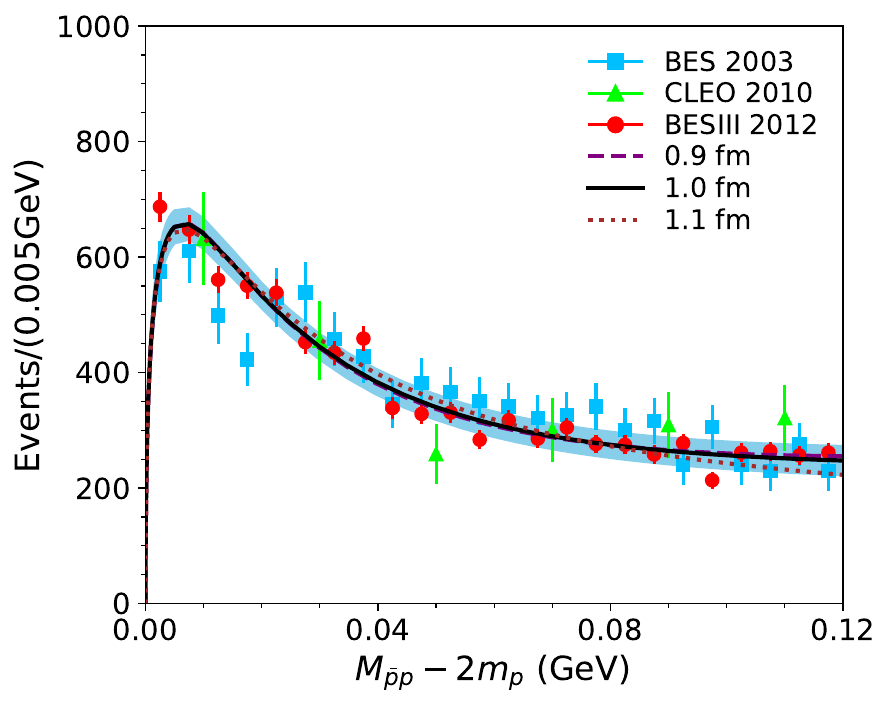}
    \includegraphics[width=0.9\linewidth]{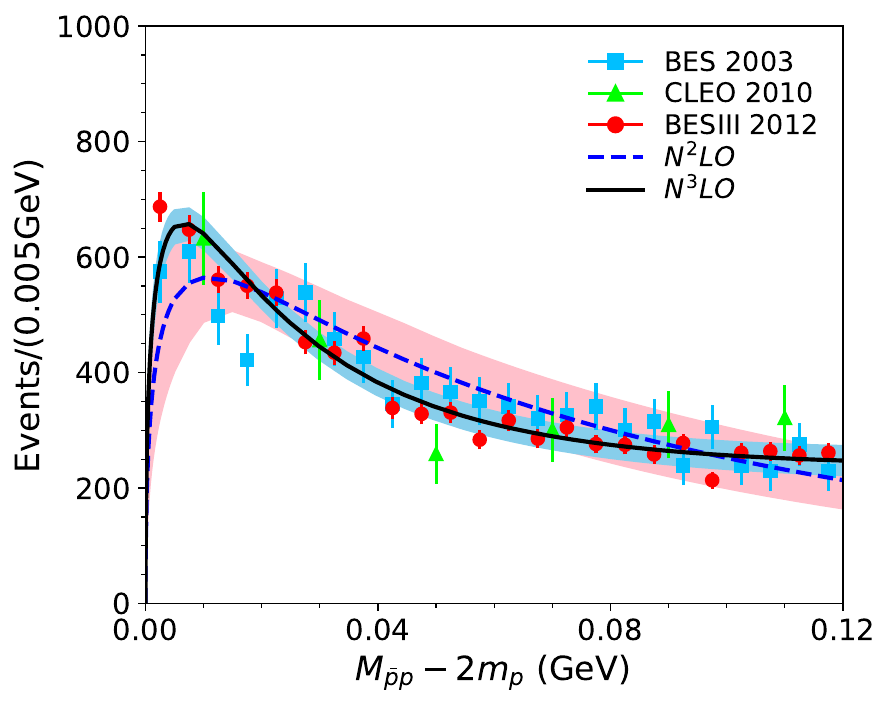}
    \caption{Fit to the $\Jp\to\gamma p\bar{p}$ decay rate. At the top, they are of N$^3$LO with cutoffs $R$=0.9, 1.0, and 1.1 fm. The sky blue band is the error estimation of the results with $R$=1.0 fm. The data are from BESIII~\cite{BESIII:2011aa}, BES~\cite{BES:2003aic} and CLEO~\cite{CLEO:2010fre}. At the bottom, the graphs are the results for N$^2$LO and N$^3$LO, with cutoff $R=1.0$~fm, respectively. The pink and sky blue bands are the uncertainties for N$^2$LO and N$^3$LO, respectively.}
    \label{decay3}
\end{figure}

With the $N\bar{N}$ scattering amplitude, one can obtain the $\Jp\to \gamma N\bar{N}$ decaying amplitude according to  Eq.~(\ref{eq:DWBA}). Here we set $C_{\Jp\to\gamma p\bar{p}}=0$ as a constant transition amplitude $A^0_{\Jp\to\gamma N\bar{N}}$ is enough to describe the data well. Also, we fix $\tilde{C}_{\Jp\to\gamma p\bar{p}}=1$ due to the overall normalization factor for the events data.
Notice that the phase shifts and inelasticity of $I$=1 $^1S_0$ partial wave and the invariant mass spectra from BESIII~\cite{BESIII:2011aa}, BES~\cite{BES:2003aic} and CLEO~\cite{CLEO:2010fre} are fitted simultaneously. The fitting results of the decay rate of the process of $\Jp\to\gamma p\bar{p}$ are shown in Fig.\ref{decay3}. 
The top graphs of Fig.~\ref{decay3} are those of N$^3$LO with different cutoffs: 
The purple dashed, black solid, and brown dotted curves are for cutoffs $R$=0.9, 1.0, and 1.1 fm, respectively. All curves with different cutoffs are consistent with the experimental data. The results with $R$=0.9 and 1.0 fm are somewhat similar to each other, and they have only a slight difference with the result of $R$=1.1 fm. 
The comparison between the decay rates of N$^2$LO and N$^3$LO are given in the bottom graphs of Fig.~\ref{decay3}, shown as the dashed blue and solid black curves. 
The error bands of the results with $R$=1.0 fm are given by the pink and sky blue bands for N$^2$LO and N$^3$LO, respectively. The estimation is based on a Bayesian method~\cite{Epelbaum2015,Dai:2017ont}. In the next sections, we will also use this method to estimate the uncertainties. 
Though the fit of N$^2$LO is not as good as that of N$^3$LO, especially in the energy region around the threshold, 
both are compatible with the experimental measurements. 

As observed by the experiments, there is a clear threshold enhancement near the $p\bar{p}$ threshold. This inspires people to believe that there is a baryonium state related to the $N\bar{N}$,  e.g., the $X(1835)$. In our analysis, such a state is not found in the origin $N\bar{N}$ scattering amplitudes, neither in $I=0$ nor in $I=1$ $^1S_0$ waves. Nonetheless, with the modest modified $I=1$ $^1S_0$ partial wave, such a bound state can be found below the $p\bar{p}$ threshold. For instance, the pole location for $R=1.0$~fm is $E_B=-49.8-i49.2$ MeV of N$^3$LO, and $E_B=-1.0-i100.2$ MeV of N$^2$LO, which are compatible with those found in Refs.~\cite{Kang:2015yka,Dai:2018tlc}.
Nevertheless, the width ($\Gamma=2~{\rm Im}E_B$) of the state is large, and thus it is not easy to conclude that the anomalous structure around $N\bar{N}$ threshold of other processes is purely caused by the $X(1835)$. 
It should be stressed that the pole found here may not be suitable to be treated as evidence of confirming the baryonium origin of the $X(1835)$\footnote{Indeed, in some other processes, there have been observed (baryon-anti-baryon) threshold enhancements. However, it is not necessary to have a baryonium state there \cite{Dai:2017fwx,Haidenbauer:2020wyp}.}, as the data above the $N\bar{N}$ or $B\bar{B}$ threshold could determine the amplitudes below threshold unreliably.

\subsection{Analysis of \texorpdfstring{$J/\psi\to \gamma \eta'\pi^+\pi^-$}{}}
As mentioned above, to determine the $\Jp\to\gamma \eta'\pi^+\pi^-$ decay amplitude, one needs to know the amplitudes of two processes, $\Jp\to\gamma N\bar{N}$ and  $ N\bar{N}\to\eta'\pi^+\pi^-$. The former decaying amplitude has been given in the previous subsection, and the latter is calculated out by the first equation of Eq.~(\ref{eq:DWBA}). 
Notice that only isoscalar $^1 S_0$ wave is needed. 
The unknown couplings $\tilde{C}_{N\bar{N}\to\eta'\pi\pi}$ ($C_{N\bar{N}\to\eta'\pi\pi}$ is set to be zero) is fixed by fitting to the cross section, $\sigma(p\bar{p}\to\eta'\pi^+\pi^-)=2.23$ mb, which is obtained by multiplying the measured branching ratio BR($p\bar{p}\to\eta'\pi^+\pi^-$)=0.626$\%$~\cite{Amsler:2004rd} and the total annihilation cross section of $p\bar{p}$~\cite{OBELIX:1996pze}.
With these two amplitudes, we can now use DWBA to get the decaying amplitude of $\Jp\to\gamma \eta'\pi^+\pi^-$. 
Note that the parameters $\tilde{C}_{\Jp\to\gamma N\bar{N}}$, $\tilde{C}_{N\bar{N}\to\eta'\pi\pi}$ and normalization factors for the $M_{\eta'\pi^+\pi^-}$ invariant mass spectra are multiplied together. Therefore,  we can set $\tilde{C}_{\Jp\to\gamma N\bar{N}}=1$ again. 
Finally, one only needs to determine three parameters by fitting to the $\Jp\to\gamma\eta'\pi\pi$ invariant mass spectra, i.e., $\tilde{C}_{\Jp\to\gamma\eta'\pi\pi}$, $C_{\Jp\to\gamma\eta'\pi\pi}$ and a normalization factor. Notice that $\tilde{C}_{\Jp\to\gamma\eta'\pi\pi}$ and $C_{\Jp\to\gamma\eta'\pi\pi}$ can be complex numbers, and hence there are four degrees of freedom for them. 
In the previous work \cite{Dai:2018tlc}, these two parameters are taken as real numbers. 
However, in this analysis, for cutoffs R=1.0 and 1.1 fm, the complex parameters would give a better quality of the fits. The parameters of our solutions are shown in Table \ref{Tab1}.
\begin{table}[htp]
    \centering
    \begin{tabular}{c|c|ccc}
    \hline\hline
                                                       & N$^2$LO & \multicolumn{3}{c}{N$^3$LO}\\\hline
     $R$~(fm)                                                        & 1.0      &  0.9     & 1.0      & 1.1   \\\hline
    $\tilde{C}_{\bar{p}p\to\eta'\pi\pi}$~(GeV$^{-2}$)                & 0.0075   & 0.0069   & 0.0072   & 0.0075\\
    \multirow{2}*{$\tilde{C}_{\Jp\to\gamma\eta'\pi\pi}$~(GeV$^{-2}$)}&-0.2498   &-0.1072   &-0.1561   &-0.2283\\
                                   ~                                 &+0.0105$i$&-0.0844$i$&-0.0342$i$&+0.0177$i$\\
    \multirow{2}*{$C_{\Jp\to\gamma\eta'\pi\pi}$~(GeV$^{-3}$)}        & 0.0881   &  0.0360  &0.0538    &0.0830\\
                                   ~                                 &+0.0101$i$&+0.0462$i$&+0.0249$i$&+0.0025$i$\\
    \hline\hline
    \end{tabular}
\caption{Values of the real parameter  $\tilde{C}_{\bar{p}p\to\eta'\pi\pi}$ and complex ones $\tilde{C}_{\Jp\to\gamma\eta'\pi\pi}$ and $C_{\Jp\to\gamma\eta'\pi\pi}$
appearing in Eq.~(\ref{AV}). For the complex parameters, 
the first line corresponds to the real part, and the second line to the
imaginary part. All these parameters are in units of $10^4$.}
\label{Tab1}
\end{table}

\begin{figure}[htp]
    \centering
    \includegraphics[width=0.9\linewidth]{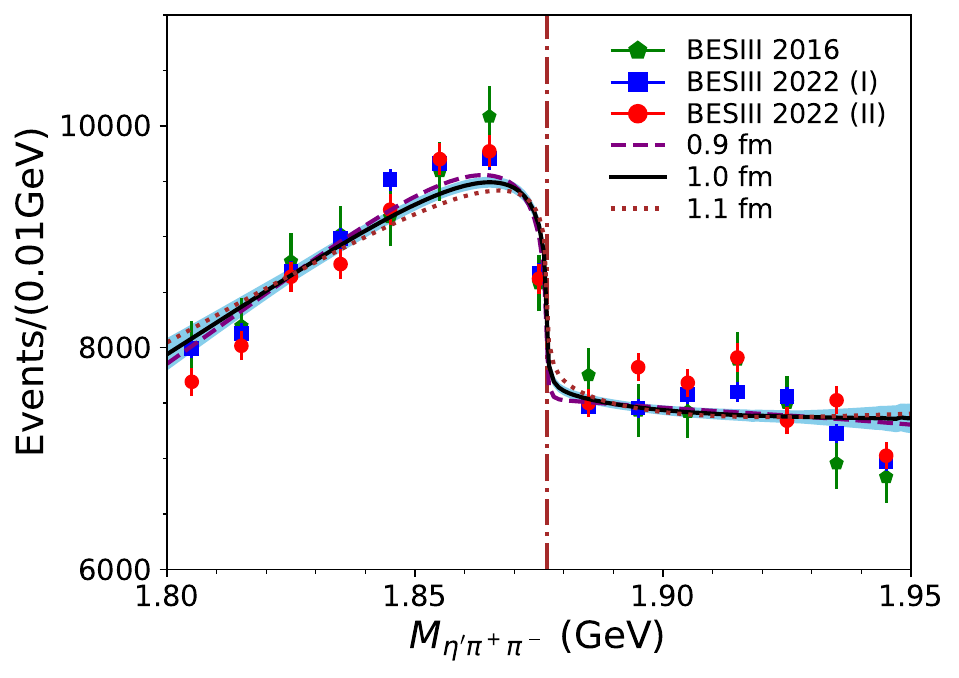}
    \includegraphics[width=0.9\linewidth]{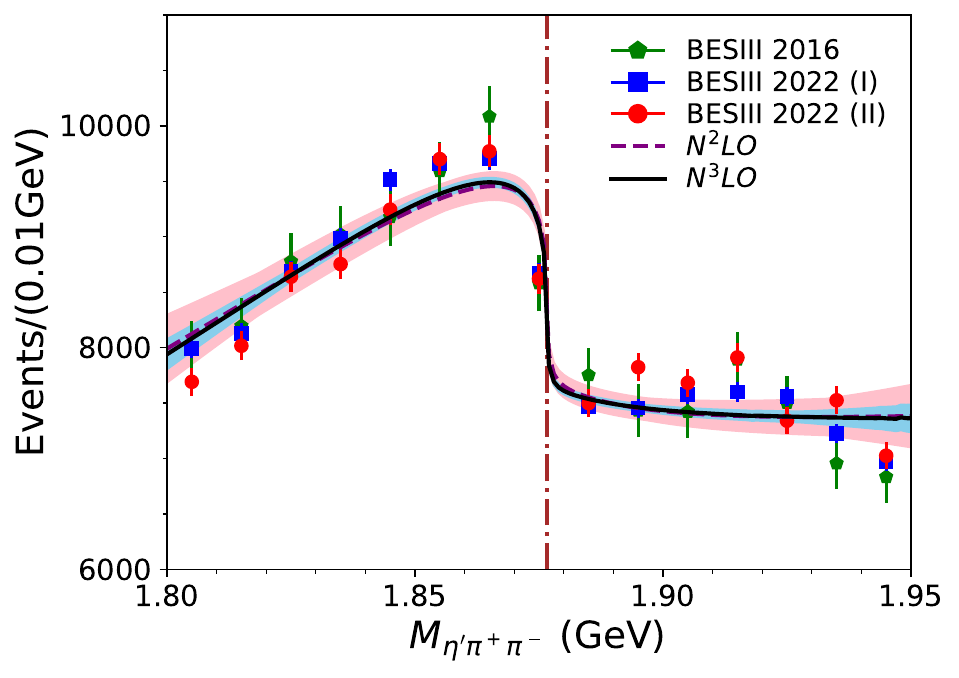}
    \caption{Fit to the $\Jp\to\gamma\eta'\pi^+\pi^-$ decay rate. The top graphs are of N$^3$LO with cutoffs $R=$0.9, 1.0, and 1.1 fm. The bottom graphs are for N$^2$LO and N$^3$LO, with cutoff $R$=1.0 fm. The datasets labeled "BESIII 2022 (I) and "BESIII 2022 (II) are from Ref.\cite{Ablikim2022} and the one labeled "BESIII 2016" is from Ref.~\cite{Ablikim2016}. The pink and sky blue bands are the error estimation of the results for N$^2$LO and N$^3$LO, respectively, with $R$=1.0 fm. The vertical line is the $p\bar{p}$ threshold. }
    \label{jpsdecay1}
   \end{figure}

Recently, BESIII collaboration performed a new measurement on the invariant mass spectra of the decay rate of $\Jp\to \gamma \eta'\pi^+\pi^-$ \cite{Ablikim2022}. 
In this new measurement, two ways to collect the events of the final state $\eta'$ are considered: one is from $\eta'\to\gamma\pi^+\pi^-$, and the other is from $\eta'\to\pi^+\pi^-\eta$. Correspondingly, there are two different datasets, and we label them as BESIII 2022 (I) and (II), respectively. These new datasets have smaller errors than that of the previous measurement\cite{Ablikim2016}, and thus it is necessary to include them in the present analysis. 
The fitting results are shown in Fig.~\ref{jpsdecay1}. 
The fitting results of N$^3$LO with cutoffs $R=$0.9, 1.0, and 1.1 fm are shown in the top graphs of Fig.~\ref{jpsdecay1}, denoted by dashed purple, solid black, and dotted brown curves, respectively. It can be found that all our solutions are consistent with the datasets. The difference between the results of different cutoffs is relatively tiny, revealing that the solutions are somewhat cutoff independent. 
The error bands of $R=1.0$~fm, both for  N$^2$LO and N$^3$LO,  are shown by the pink and sky blue bands as plotted in Fig.~\ref{jpsdecay1}. The error estimation method is again from the Bayesian method as Ref.~\cite{Epelbaum2015,Dai:2017ont}. Interestingly, the solution of N$^2$LO has almost the same quality as that of N$^3$LO. This is because we have let the parameters of the transition amplitude $A^0_{\Jp\to\gamma\eta'\pi^+\pi^-}$, $\tilde{C}_{\Jp\to\gamma\eta'\pi\pi}$ and $C_{\Jp\to\gamma\eta'\pi\pi}$ be complex, and thus the degrees of freedom of the fit have been increased. Also, notice that the results of $\Jp\to\gamma p\bar{p}$ at N$^2$LO are worse than that at N$^3$LO. 
Nevertheless, the mechanism of how the structure around $p\bar{p}$ threshold is generated would be similar to that of the previous paper \cite{Dai:2018tlc}. See discussions below.

\begin{figure}[htp]
    \centering
    \includegraphics[width=0.95\linewidth]{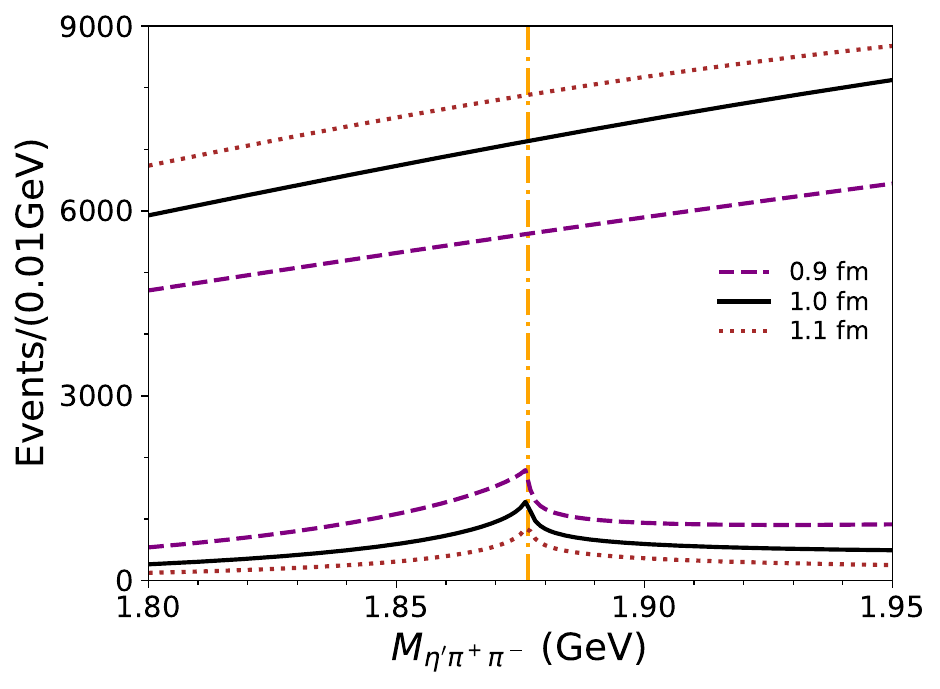}
    \caption{The individual contribution to $M_{\eta'\pi^+\pi^-}$ invariant mass spectra of $\Jp\to\gamma\eta'\pi^+\pi^-$, either from the background (upper lines) or from the $N\bar{N}$ rescattering (lower lines with cusplike behavior around $p\bar{p}$ threshold). The contributions are of N$^3$LO for cutoffs $R=$0.9, 1.0, and 1.1 fm, respectively. }
    \label{jpsidecay2}
\end{figure}

To study the structure around $p\bar{p}$ threshold in the $M_{\eta'\pi^+\pi^-}$ invariant mass spectra of $\Jp\to\gamma \eta'\pi^+\pi^-$, we separate two contributions to the $M_{\eta'\pi^+\pi^-}$ spectra: one is from $N\bar{N}$ rescattering ($\Jp\to\gamma \bar{N}N\to\gamma \eta'\pi^+\pi^-$), and the other is the \lq background, i.e., the contribution from the transition amplitude $A^0_{\Jp\to\gamma\eta'\pi\pi}$. 
The individual contributions of these two parts are calculated by Eq.~(\ref{eq:DWBA}), with one part switched on while the other switched off.  
The results are shown in Fig.\ref{jpsidecay2}. 

As can be seen, the backgrounds are pretty smooth and more prominent, while the $N\bar{N}$ rescattering term contributes little except for a cusplike structure around the $p\bar{p}$ threshold.
This confirms our conclusion,  i.e., the structure around the $p\bar{p}$ threshold in the $M_{\eta'\pi^+\pi^-}$ invariant mass spectra of $\Jp\to\gamma \eta'\pi^+\pi^-$ should be generated by the interference between the background and the $N\bar{N}$ rescattering. 
On the other hand, there is no pole ($N\bar{N}$ bound state or resonance) found in the  $I=0$ $^1S_0$ partial wave of $N\bar{N}$ scattering amplitude for any of these cutoffs, neither in the N$^2$LO case nor in the  N$^3$LO case\footnote{We are aware that in some other models \cite{Dedonder:2009bk,Milstein:2017dqp,Dedonder:2018ulo}, they find a near-threshold pole in the $N\bar{N}$ $I=0$ $^1S_0$ partial wave.   }. 
Hence, our analysis indicates that there is no need for a $N\bar{N}$ bound state to quantitatively describe the structure around the $p\bar{p}$ threshold in $M_{\eta'\pi^+\pi^-}$ invariant mass spectra of $\Jp\to\gamma \eta'\pi^+\pi^-$.

\subsection{Analysis of $\Jp\to\gamma 3(\pi^+\pi^-)$}
As discussed above, the decaying amplitude of $\Jp\to\gamma 3(\pi^+\pi^-)$ is constructed by DWBA with the amplitudes of $\Jp\to\gamma N \bar{N}$ and $N\bar{N}\to 3(\pi^+\pi^-)$. The former has been given in the previous sections, and the latter is fixed by fitting the relevant cross section. 
In Ref.~\cite{Sai:1982dv}, the $p\bar{p}\to 3(\pi^+\pi^-)$ cross section is measured in the momentum range $p_{\mathrm{lab}}$=374-680 MeV$/c$. In the low energy region, one can obtain $\sigma(p\bar{p}\to 3(\pi^+\pi^-))$=7.52$\pm$0.89 mb at $p_{\mathrm{lab}}$=106.6 MeV$/c$ by combining the branching ratio BR($p\bar{p}\to 3(\pi^+\pi^-)$)=$21.0\pm2.5$\%~\cite{Klempt:2005pp} and the total $p\bar{p}$ annihilation cross section~\cite{OBELIX:1996pze}.
With the amplitudes of $\Jp\to\gamma N \bar{N}$ and $N\bar{N}\to 3(\pi^+\pi^-)$,  we can construct the decay amplitude of $\Jp\to\gamma 3(\pi^+\pi^-)$ according to Eq.~(\ref{eq:DWBA}). The unknown couplings of the Born term of $J/\psi\to\gamma 3(\pi^+\pi^-)$ and the potential of $N\bar{N}\to 3(\pi^+\pi^-)$ will be fixed by combination fitting the cross section of $p\bar{p}\to 3(\pi^+\pi^-)$ and the invariant mass spectrum of $\Jp\to\gamma 3(\pi^+\pi^-)$ \cite{BESIII:2013sbm}. 

The unknown parameters are $\tilde{C}_{\bar{p}p\to 3(\pi^+\pi^-)}$, $C_{\bar{p}p\to 3(\pi^+\pi^-)}$, $\tilde{C}_{\Jp\to\gamma 3(\pi^+\pi^-)}$ and $C_{\Jp\to\gamma 3(\pi^+\pi^-)}$, see Eq.~(\ref{AV}). Note that only the isoscalar $^1S_0$ partial wave is taken into account according to ignoring higher partial waves. For $\tilde{C}_{\Jp\to\gamma 3(\pi^+\pi^-)}$ and $C_{\Jp\to\gamma 3(\pi^+\pi^-)}$, they can be complex numbers, resulting in four degrees of freedom. For $C_{p\bar{p}\to 3(\pi^+\pi^-)}$ that corresponds to momentum dependence of the potential, we include it and find a better fit. Besides, one also needs a normalization factor for the decay rates of $\Jp\to\gamma 3(\pi^+\pi^-)$ in the full range. The values of parameters of our fit are given in Table~\ref{Tab2}.
\begin{table}[htp]
	\centering
	\begin{tabular}{c|c|ccc}
		\hline\hline
		& N$^2$LO & \multicolumn{3}{c}{N$^3$LO}\\\hline
		$R$~(fm)                                                            & 1.0      &  0.9     & 1.0      & 1.1   \\\hline
		$\tilde{C}_{\bar{p}p\to 3(\pi^+\pi^-)}$~(GeV$^{-2}$)                &0.0124    &0.0108    &0.0113    &0.0117\\
		$C_{\bar{p}p\to 3(\pi^+\pi^-)}$~(GeV$^{-2}$)                        &0.1246    &0.0980    &0.1245    &0.1555\\
		\multirow{2}*{$\tilde{C}_{\Jp\to\gamma 3(\pi^+\pi^-)}$~(GeV$^{-2}$)}&0.3578    &0.1491    &0.1453    &0.2015\\
		~                                                                   &-0.1786$i$&-0.0477$i$&-0.1257$i$&-0.1080$i$\\
		\multirow{2}*{$C_{\Jp\to\gamma 3(\pi^+\pi^-)}$~(GeV$^{-3}$)}        &-0.2009   &-0.0350   &-0.0522   &-0.1021\\
		~                                                                   &+0.0592$i$&+0.0149$i$&+0.0447$i$&0.0291$i$\\
		\hline\hline
	\end{tabular}
	\caption{As in Table II but for the real parameters $\tilde{C}_{\bar{p}p\to3(\pi^+\pi^-)}$, $C_{\bar{p}p\to3(\pi^+\pi^-)}$ and the complex ones $\tilde{C}_{J/\psi\to\gamma 3(\pi^+\pi^-)}$ and $C_{J/\psi\to\gamma 3(\pi^+\pi^-)}$.
	\label{Tab2} }
\end{table}

\begin{figure}[htp]
	\centering
	\includegraphics[width=0.9\linewidth]{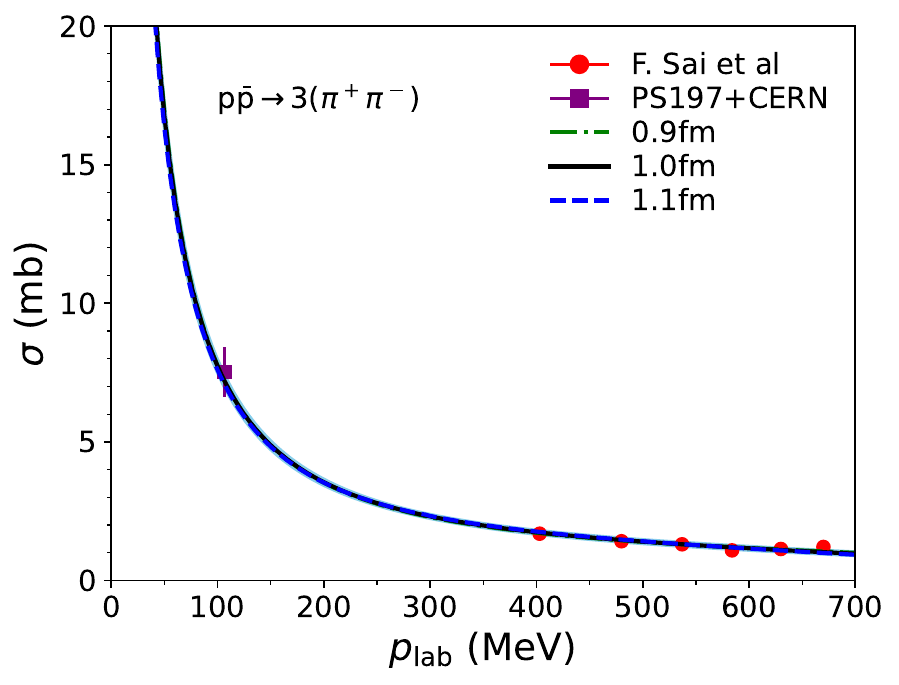}
	\includegraphics[width=0.9\linewidth]{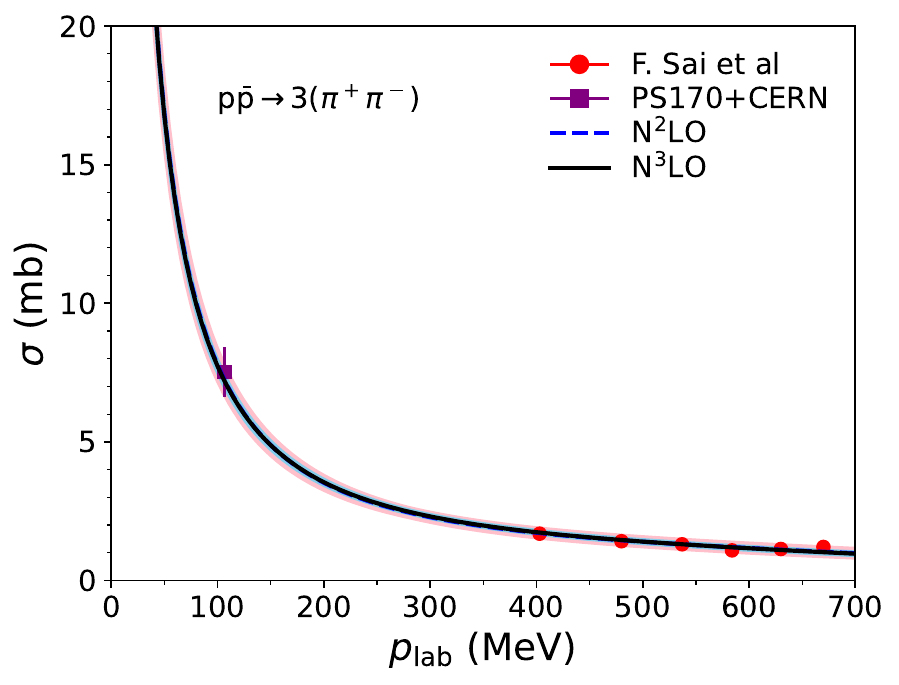}
	\caption{Fit to the $p\bar{p}\to 3(\pi^+\pi^-)$ cross section. The top graphs are of N$^3$LO with cutoffs $R=$0.9, 1.0, and 1.1 fm. The bottom graphs are for N$^2$LO and N$^3$LO, with cutoff $R$=1.0 fm. The pink and sky blue bands are the uncertainties for the solutions of N$^2$LO and N$^3$LO, respectively. Here the \lq F.Sai et al.' data
    are from Ref.~\cite{Sai:1982dv}. The data point labeled as \lq PS197 + CERN' is taken from Refs.~\cite{Klempt:2005pp,OBELIX:1996pze}.  }
	\label{jpsidecay3}
\end{figure}

The results of fitting to the cross section of $p\bar{p}\to 3(\pi^+\pi^-)$ are shown in Fig.~\ref{jpsidecay3}. 
The top graphs show the results of N$^3$LO with different cutoffs, $R$=0.9, 1.0, and 1.1 fm. Our solutions are consistent with the data. Indeed, there is almost no difference between the solutions with different cutoffs. The results for N$^2$LO and N$^3$LO, with the cutoff $R$=1.0 fm, are shown in the bottom graphs. 
As can be found, they are consistent with the data, and the one of N$^2$LO (the dashed blue line and pink band) overlaps with that of N$^3$LO (the solid black line and sky blue band). 
From this aspect, one can conclude that the higher order correction on $N\bar{N}$ rescattering at N$^3$LO has little influence on the scattering of $p\bar{p}\to 3(\pi^+\pi^-)$. The reason may be that there are only a few data points for this process, so the N$^2$LO description on the $N\bar{N}$ rescattering is enough.

\begin{figure}[htp]
	\centering
	\includegraphics[width=0.9\linewidth]{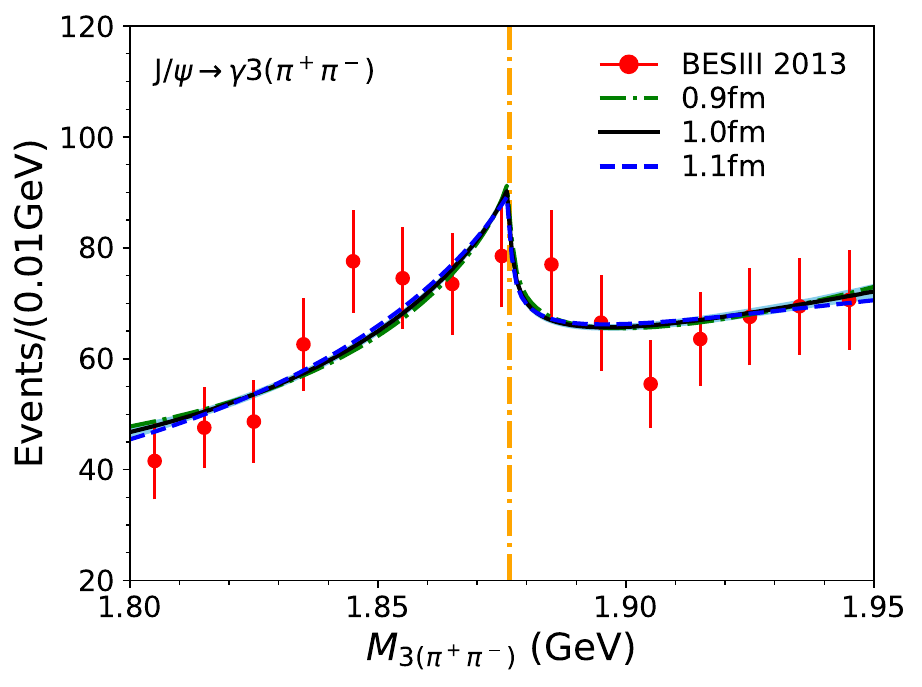}
	\includegraphics[width=0.9\linewidth]{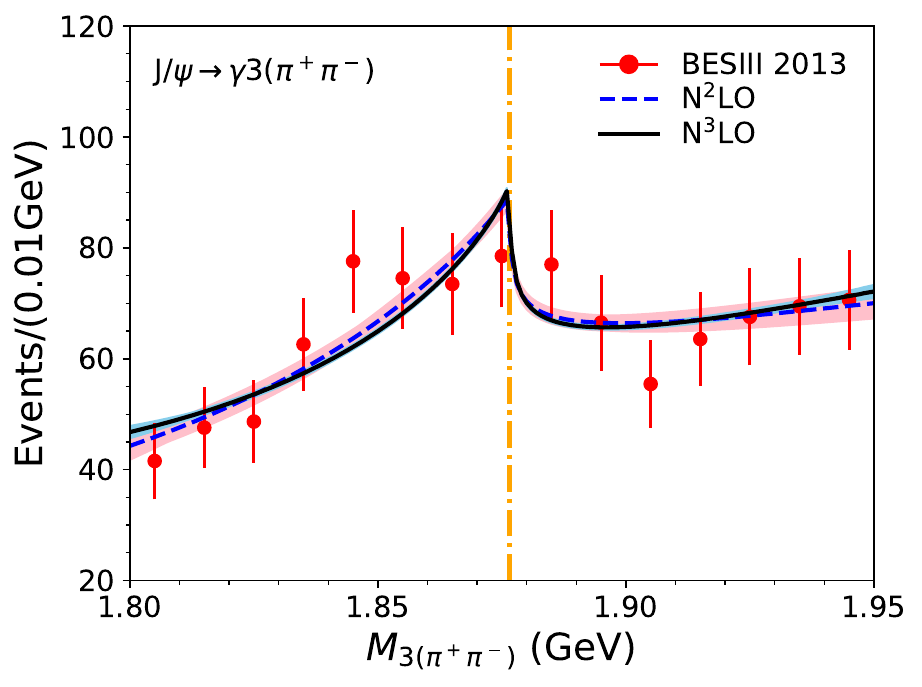}
	\caption{As in Fig.~\ref{jpsdecay1} but for the $J/\psi\to\gamma 3(\pi^+\pi^-)$.
    Here the BESIII data are from Ref. \cite{BESIII:2013sbm}.}
	\label{jpsidecay4}
\end{figure}

The results of fitting to the $\Jp\to\gamma 3(\pi^+\pi^-)$ decay rates are shown in Fig.~\ref{jpsidecay4}. 
In the top graphs, the dash-dotted green, solid black, and dashed blue lines are the solutions with cutoffs $R$=0.9, 1.0, and 1.1 fm of N$^3$LO, respectively. 
In the bottom graphs, the dashed blue and black solid lines are for N$^2$LO and N$^3$LO with cutoff $R$=1.0 fm. Notice that we only focus on the energy region around the $p\bar{p}$ threshold. 
One would notice that our solutions around the $N\bar{N}$ threshold behaves like a cusp, while two data points, one around 1.84~GeV and the other around 1.90~GeV, are not fitted well. Nevertheless, taking into account the large uncertainties of the data,  our solutions are still compatible with the data of the $M_{3(\pi^+\pi^-)}$ invariant mass spectrum of $\Jp\to\gamma 3(\pi^+\pi^-)$. The difference between the solutions with different cutoffs is tiny. See the green, black and blue curves in the top graphs. This implies that our solutions of $\Jp\to\gamma 3(\pi^+\pi^-)$ are stable by varying cutoffs. The results of N$^2$LO and N$^3$LO also have slight differences. See the blue and black curves at the bottom. The reason may be that the data statistics need to be higher.

\begin{figure}[htp]
	\centering
	\includegraphics[width=0.9\linewidth]{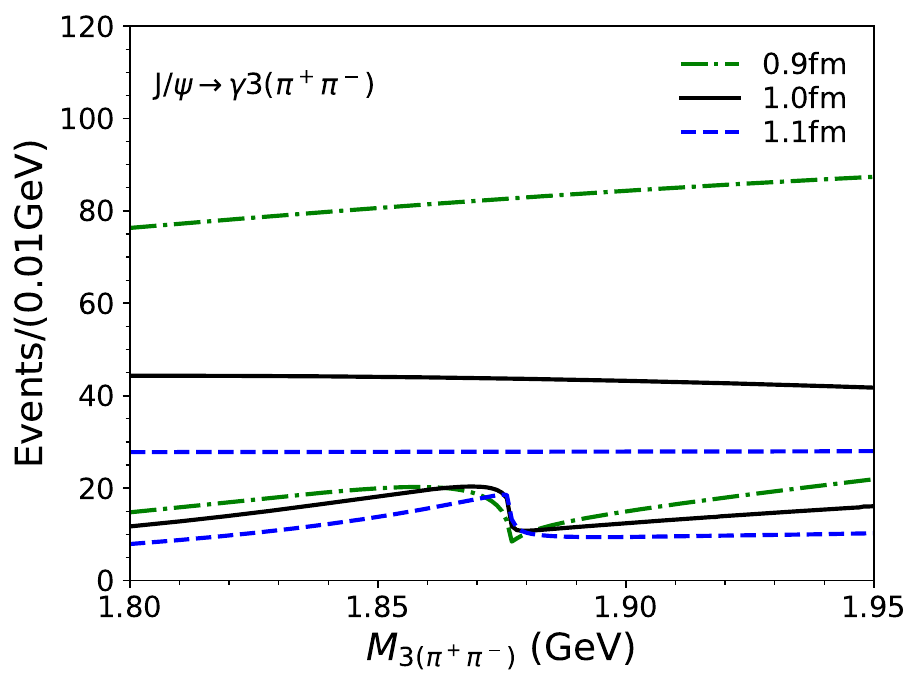}
	\caption{The individual contribution to $M_{3(\pi^+\pi^-)}$ invariant mass spectra of $\Jp\to\gamma 3(\pi^+\pi^-)$.}
	\label{jpsidecay5}
\end{figure}

Like what is done in the previous subsection, 
we present the individual contributions from the background and the $N\bar{N}$ rescattering ($\Jp\to\gamma N\bar{N}\to\gamma 3(\pi^+\pi^-)$). See Fig.~\ref{jpsidecay5}.
The dash-dotted green, solid black, and dashed blue lines are for cutoffs $R$=0.9, 1.0, and 1.1 fm, respectively. The contributions from the background are smooth, and the structure around the $p\bar{p}$ threshold is caused by the $N\bar{N}$ rescattering. This supports that the structure of $M_{3(\pi^+\pi^-)}$ invariant mass spectrum around the $p\bar{p}$ threshold should be originated from the $N\bar{N}$ threshold behavior.

\subsection{Analysis of $\Jp\to\gamma K_S^0 K_S^0\eta$}
The amplitude of $\Jp\to\gamma K_S^0 K_S^0\eta$ is constructed through a two-step process of $\Jp\to\gamma N\bar{N}\to\gamma K_S^0 K_S^0\eta$. The amplitude of $p\bar{p}\to K_S^0 K_S^0\eta$ is fixed by the cross section, which is obtained from the branching ratio BR($p\bar{p}\to K_S^0 K_S^0\eta$)=(0.25$\pm$0.04)$\times 10^{-3}$\% \cite{Barash:1967dts} and the total annihilation cross section of $p\bar{p}$~\cite{OBELIX:1996pze}. One has $\sigma(p\bar{p}\to K_S^0 K_S^0\eta)=0.89\pm0.14$ $\mu$b at $p_{\mathrm{lab}}=106.6$ MeV$/c$. Combined the amplitudes of $\Jp\to\gamma K_S^0 K_S^0\eta$ and $p\bar{p}\to K_S^0 K_S^0\eta$ through DWBA, one can fix the amplitude of $\Jp\to\gamma K_S^0 K_S^0\eta$ by fitting to the $M_{K_S^0K_S^0\eta}$ invariant mass spectrum. Notice that in Ref.~\cite{BESIII:2015xco}, the structure around the $X$(1835) become more prominent by requiring $M_{K_S^0K_S^0}<1.1$ GeV$/c^2$. Correspondingly, we set the upper limit $t_1^+=(1.1~\mathrm{GeV}/c^2)^2$ in Eq.~\eqref{eq:dGdQ} for the decay rate of $J/\psi\to\gamma K_S^0 K_S^0\eta$. 
Notice that the $f_0(980)\eta$ channel is quite far away from $N\bar{N}$ threshold, and it will only supply a background in the energy region concerned. 

The unknown parameters are given in Eq.~\eqref{AV}, i.e., $\tilde{C}_{\Jp\to\gamma K_S^0 K_S^0\eta}$, $C_{\Jp\to\gamma K_S^0 K_S^0\eta}$, $\tilde{C}_{\bar{p}p\to K_S^0 K_S^0\eta}$ and $C_{\bar{p}p\to K_S^0 K_S^0\eta}$. Since there is only one data point for the $p\bar{p}\to K_S^0 K_S^0\eta$ cross section, we set $C_{\bar{p}p\to K_S^0 K_S^0\eta}$ to be zero and then $\tilde{C}_{\bar{p}p\to K_S^0 K_S^0\eta}$ can be fixed. For $\tilde{C}_{\Jp\to\gamma K_S^0 K_S^0\eta}$ and $C_{\Jp\to\gamma K_S^0 K_S^0\eta}$, they could be complex numbers, but in practice, we can set them as real numbers and the fit is in high quality. Thus there are only two degrees of freedom. In addition, an overall normalization factor to the data of  $M_{K_S^0 K_S^0\eta}$ invariant mass spectrum in the $J/\psi$ radiative decay is needed. 
The values of these parameters are shown in Table \ref{Tab3}.
\begin{table}[htp]
	\centering
	\begin{tabular}{c|c|ccc}
		\hline\hline
		& N$^2$LO & \multicolumn{3}{c}{N$^3$LO}\\\hline
		$R$~(fm)                                             & 1.0   &  0.9  & 1.0   & 1.1   \\\hline
		$\tilde{C}_{\bar{p}p\to K_S^0 K_S^0\eta}$~(GeV$^{-2}$) & 0.4659& 0.4514& 0.4573& 0.4607\\
		$\tilde{C}_{\Jp\to\gamma K_S^0 K_S^0\eta}$~(GeV$^{-2}$)& 6.8187& 0.9990& 4.4638& 6.1870\\
		$C_{\Jp\to\gamma K_S^0 K_S^0\eta}$~(GeV$^{-3}$)        &-4.2018&-0.5561&-2.6581&-3.7904\\
		\hline\hline
	\end{tabular}
	\caption{The values of the real parameters, $\tilde{C}_{\bar{p}p\to K_S^0 K_S^0\eta}$, $\tilde{C}_{\Jp\to\gamma K_S^0 K_S^0\eta}$ and $C_{\Jp\to\gamma K_S^0 K_S^0\eta}$ as given in 
    Eq.~\eqref{AV}. All parameters are in units of $10$. \label{Tab3} }
\end{table}
One would notice that $\tilde{C}_{\bar{p}p\to K_S^0 K_S^0\eta}$ are much smaller than that of $\tilde{C}_{\bar{p}p\to \eta'\pi^+\pi^-}$, $\tilde{C}_{\bar{p}p\to \eta'\pi^+\pi^-}$ and $\tilde{C}_{\bar{p}p\to 3(\pi^+\pi^-)}$. This is because the cross section of $p\bar{p}\to K_S^0 K_S^0\eta$ is almost three orders smaller than that of $p\bar{p}\to \eta'\pi^+\pi^-$ and $p\bar{p}\to 3(\pi^+\pi^-)$.
For $\tilde{C}_{J/\psi\to\gamma K_S^0 K_S^0\eta}$ and $C_{J/\psi\to\gamma K_S^0 K_S^0\eta}$,  it is difficult to discuss why its magnitude is smaller than those of $\tilde{C}_{J/\psi\to 3(\pi^+\pi^-)}$, $C_{J/\psi\to 3(\pi^+\pi^-)}$, $\tilde{C}_{J/\psi\to \eta'\pi^+\pi^-}$ and $C_{J/\psi\to \eta'\pi^+\pi^-}$, because these transition potentials interfere with the $N\bar{N}$ rescattering parts. Their magnitudes are fixed by fitting to the invariant mass spectra. Notice that the whole amplitudes should also multiply an overall normalization factor to fit the events data. 

\begin{figure}[htp]
	\centering
	\includegraphics[width=0.9\linewidth]{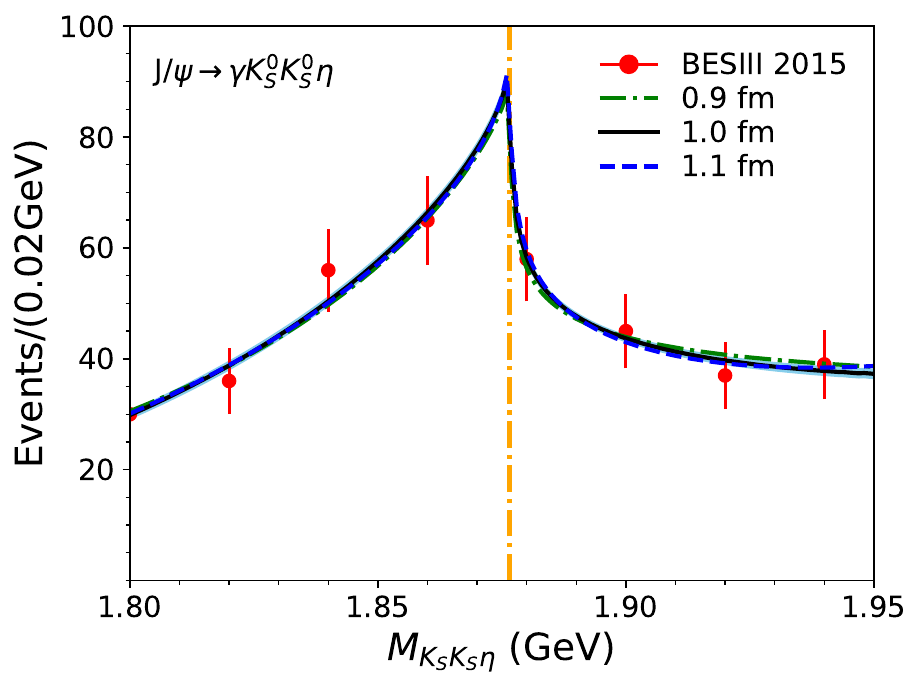}
	\includegraphics[width=0.9\linewidth]{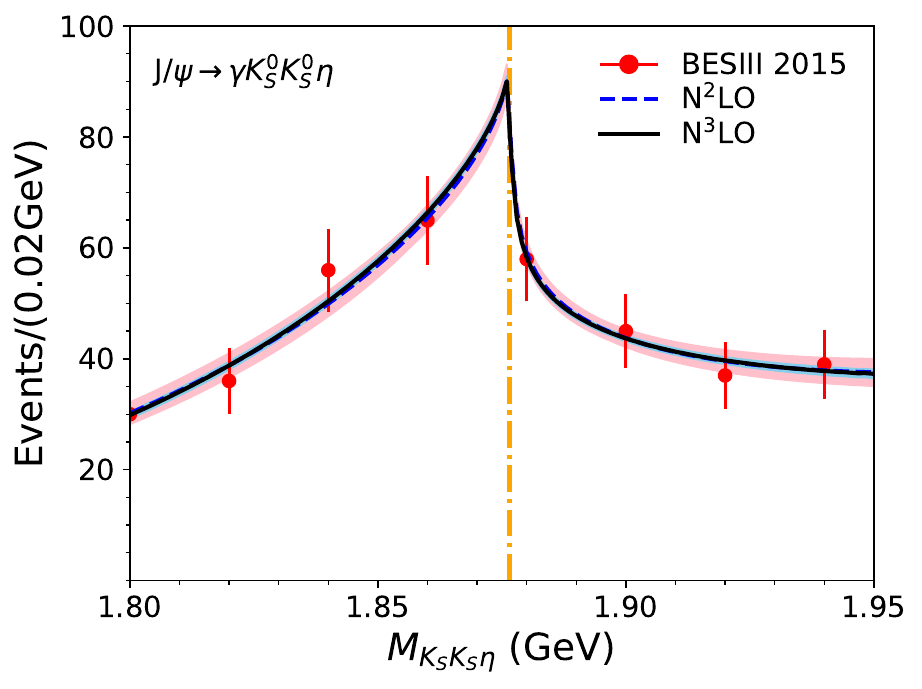}
	\caption{As in  Fig.~\ref{jpsdecay1} but for $J/\psi\to\gamma K_S^0K_S^0\eta$. Here the data is from BESIII \cite{BESIII:2015xco}.  }
	\label{jpsidecay6}
\end{figure}

Our solutions fitting to the $M_{K_S^0K_S^0\eta}$ invariant mass spectrum are shown in Fig.~\ref{jpsidecay6}.
As can be found, our solutions fit the experimental data rather well. However, it should be noted that there are only 8 data points in the energy region of [1.80-1.95] GeV that we focus on.  
The top graphs in Fig.~\ref{jpsidecay6} are the results of N$^3$LO. The dash-dotted green, solid black, and dashed blue lines are for cutoffs $R=$0.9, 1.0, and 1.1 fm, respectively. These curves of different cutoffs almost overlap, implying that the results are not sensitive to the cutoffs. The bottom graphs are the results of N$^2$LO (dashed blue) and N$^3$LO (black solid) with cutoff $R$=1.0 fm. 
The results of N$^2$LO and N$^3$LO overlap with each other, and this is again caused by lacking enough statistics on the data. 
It would be rather helpful if there are future measurements on the invariant mass spectrum of $\Jp\to\gamma K_S^0 K_S^0 \eta$, especially those around the $p\bar{p}$ threshold to get apparent structure. 

\begin{figure}[htp]
	\centering
	\includegraphics[width=0.9\linewidth]{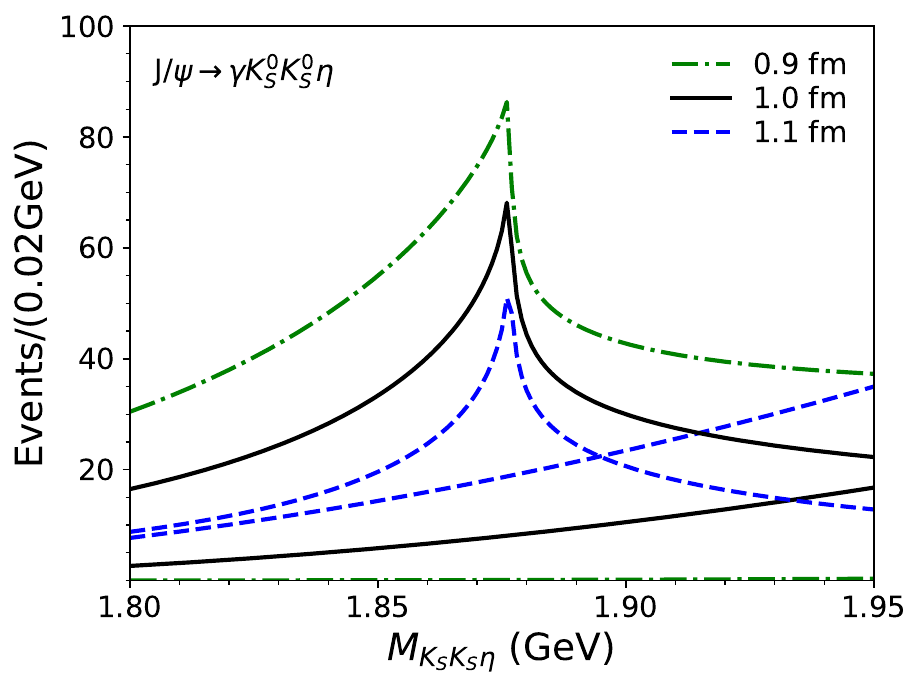}
	\caption{The individual contribution to $M_{K_S^0 K_S^0\eta}$ invariant mass spectra of $\Jp\to\gamma K_S^0 K_S^0\eta$.}
	\label{jpsidecay7}
\end{figure}
The individual contributions from background and $N\bar{N}$ rescattering to $M_{K_S^0 K_S^0\eta}$ invariant mass spectra of $\Jp\to\gamma K_S^0 K_S^0\eta$ are shown Fig.~\ref{jpsidecay7}. 
The smooth lines at the bottom are from backgrounds (Born terms as given in Eq.\eqref{AV}), and the ones with cusps around the  $p\bar{p}$ threshold are from $N\bar{N}$ rescattering. 
It shows clearly that the structure around the $p\bar{p}$ threshold is caused by $N\bar{N}$ threshold behavior. Note that in some other models, the origin of the structure around $p\bar{p}$ threshold could be caused by some bound states. See e.g., Refs.~\cite{Kang:2015yka,Dedonder:2018ulo}. As a short summary, in all the processes of $J/\psi$ radiative decays, $\Jp\to\gamma\eta'\pi^+\pi^-$, $\Jp\to\gamma3(\pi^+\pi^-)$, and $\Jp\to\gamma K_S^0K_S^0\eta$, the structure around $p\bar{p}$ threshold are caused by the $N\bar{N}$ rescattering. While there is no resonance found in the isoscalar $^1S_0$ wave.

\section{Conclusion}\label{Sec:IV}
In this paper, we analyzed the $J/\psi$ radiative decays of the processes of $\Jp\to\gamma\eta'\pi^+\pi^-$, $\Jp\to\gamma K_S^0 K_S^0\eta$, $\Jp\to\gamma 3(\pi^+\pi^-)$ and $\Jp\to\gamma p\bar{p}$ with chiral effective field theory and distorted wave Born approximation. 
The latest measurements on the $M_{\eta'\pi^+\pi^-}$, $M_{3(\pi^+\pi^-)}$, and $M_{K_S^0 K_S^0\eta}$ invariant mass spectra are taken into account. Our results show that the structures around the $p\bar{p}$ threshold found in all these processes are caused by the $N\bar{N}$ threshold effect, while no bound state or resonance is found in the relevant $I=0$ $^1S_0$ partial wave. To check the reliability, we test the effects of different cutoffs and find that our conclusion is not sensitive to the cutoffs. It would be helpful to refine the analysis if there are future measurements on the invariant mass spectra around the $p\bar{p}$ threshold of the $J/\psi$ radiative decays.

\section*{Acknowledgements}
\label{Sec:V}
We thank professor X.-W. Kang for helpful discussions. This work is supported by Joint Large Scale Scientific Facility Funds of the National Natural Science Foundation of China (NSFC) and Chinese Academy of Sciences (CAS) under Contract No.U1932110, and NSFC Grant with No.~12061141006.

\bibliography{ref.bib}

\end{document}